# 1RXS J161935.7+524630: New Polar with the Varying Accretion Modes on two Magnetic Poles

**D. V. Denisenko[1*], F. Martinelli[2]**

[1] *Sternberg Astronomical Institute, Lomonosov Moscow State University, Russia*
[2] *Astronomical Center Lajatico, Italy*

**Abstract**–We report the discovery of a new cataclysmic variable DDE 32 identified with the ROSAT X-ray source 1RXS J161935.7+524630 in Draco. The variability was originally found by D. Denisenko on the digitized Palomar plates centered at the position of X-ray source. The photometric observations by F. Martinelli at Lajatico Astronomical Center in June 2015 have shown the large amplitude (nearly 2 magnitudes) variability with a period about 100.5 minutes. Using the publicly available Catalina Sky Survey data from 2005 to 2013 we have improved the value of period to 0.0697944 days. Comparison of the archival CRTS data with more recent observations from Lajatico shows the dramatic changes in the light curve shape. Instead of a single peak present in Catalina data before 2014, there were two peaks of nearly the same height during 2015. SDSS spectrum taken in June 2009 shows prominent Helium emission lines between the bright Balmer series. He II 4686 Å line has more than 30% effective width compared to $H_\beta$ line. All those features allow us to interpret 1RXS J161935.7+524630 as a magnetic cataclysmic variable (polar) with the accretion mode changing from one pole before 2014 to two poles in 2015.

## INTRODUCTION

Magnetic cataclysmic variables (polars) are the subclass of CVs with the strong magnetic fields. They are distinguished from other types of compact white dwarf and red dwarf (WD+RD) binary systems by the absence of accretion disk and by the presence of one or two hot spots on WD surface. These properties result in the following observational characteristics: large amplitude of orbital variability (up to 2 magnitudes); sharp drops of brightness caused by the disappearance of the hot spot behind the limb; irregular light curves on the long time scales; high X-ray to optical ratio; presence of ionized Helium lines in their spectra. Those features allow to distinguish the polars from other nova-like variables and dwarf novae with accretion disks who have smoother orbital light curves and smaller amplitudes of variability on the time scales of hours. Magnetic CVs are further divided into three subtypes: AM Her ("classical" polars), BY Cam (asynchronous polars) and DQ Her (intermediate polars, or IPs), see Andronov, 2007. AM type polars have the WD rotation synchronized with the orbital period, while IPs have white dwarfs rotating faster than the orbital period, with $P_{spin}$ typically less than 0.1 $P_{orb}$. Orbital periods of polars cover nearly the same wide range as dwarf novae, from 1.3 to 4 hours (0.054–0.16 days) and rarely a bit longer.

Many polars are showing significant long-term changes of the average brightness. Some AM Her systems have prolonged high states with the "excursions" to low states lasting for several months and even years. Several systems show "chameleon" changes of the light curve caused by the accretion stream switching from one magnetic pole to the other. And in addition to that, several systems can change the accretion mode from one pole to two poles at the same time.

---

* E-mail: d.v.denisenko@gmail.com

In this article we report the discovery and the following analysis of a new cataclysmic variable DDE 32 in Draco identified with the ROSAT X-ray source 1RXS J161935.7+524630. The first chapter describes how the variable was discovered. In the next part we analyze the spectrum of this variable obtained by Sloan Digital Sky Survey. The third chapter is devoted to the photometric observations, determination of orbital period and study of long term variations of the phased light curve. Then we discuss the possible interpretation of the observed properties of this variable. Our conclusion is that 1RXS J161935.7+524630 is a new magnetic cataclysmic variable (polar of AM Her type) with the accretion mode changing from one pole to two poles.

## DISCOVERY

The variability of DDE 32 was found by one of the authors (D.D.) back in December 2007 in the process of identifying X-ray sources from the ROSAT bright source catalogue 1RXS (Voges et al., 1999). The object 1RXS J161935.7+524630 (J1619+5246 for the brevity) is one of the bright ROSAT sources with the flux of (0.135±0.017) cnts/s in 0.2–10 keV band and hardness ratios HR1=–0.45±0.09, HR2=0.42±0.18. Statistical error of 1RXS position is 8". The star USNO-B1.0 1427-0327171 with the coordinates R.A. = 16 19 35.75, Dec. = +52 46 32.0, proper motions pmRA=14 mas/yr, pmDE=−6 mas/yr and magnitudes B1=20.16 R1=19.17 B2=19.06 R2=18.25 I=17.25 was found to be highly variable on the digitized Palomar plates. Figure 1 is showing the variable on three Palomar Red plates: 1954 June 28 (POSS-I), 1991 July 03 and 1993 April 25 (POSS-II). The large amplitude of variability is obvious. The magnitude range was determined to be 16.8–20:. The nature of the object was not clear at the time of discovery. The 2MASS photometry was misleading with the (J-K) color index of 2MASS J16193575+5246326 (J=16.71±0.16 H=16.16±0.19 K=15.52±0.23) being consistent with that of quasars, despite the non-zero proper motion in USNO-B1.0. The object was considered to be too red for a cataclysmic variable (J-K=1.2±0.3) and neglected for nearly four years.

## SDSS SPECTROSCOPY

In September 2011 D.D. came across the spectrum of J1619+5246 obtained with the Sloan Digital Sky Survey on MJD=54983 (2009 June 01). It is shown at Figure 2. Despite the automatic classification as a star of B6 spectral type by SDSS processing routines, the spectrum is clearly that of a cataclysmic variable. It shows Balmer emission lines from H$\alpha$ to H$\eta$ and in addition several Helium lines, including He II 4686 Å one characteristic for polars. The object has regained the attention again and was added to the list of variable stars discovered by the first author as DDE 32. See http://scan.sai.msu.ru/~denis/VarDDE.html for the color-combined finder chart of J1619+5246 and additional information.

## PHOTOMETRY

After the confirmation of CV nature of J1619+5246 from SDSS spectrum the new variable star was added to the monitoring program with the Bradford Robotic Telescope (BRT; since 2016 Autonomous Robotic Telescope, or ART) at http://www.telescope.org/. The object was in the low state on most BRT images obtained in 2011-2014, being not visible on 60-sec unfiltered exposures with the 0.35-m Schmidt–Cassegrain telescope and Galaxy camera based on FLI MicroLine CCD (typical limit about 19m). SDSS magnitudes (u=18.87 g=18.99 r=18.85 i=18.50 z=18.28) are also on the fainter side of J1619+5246 variability range.

However, in April 2015 the object has appeared brighter than 18m on two BRT images in a row, and D.D. has asked F.M. to start an observing run at Lajatico Astronomical Center in Italy http://www.astronomicalcentre.org/. The observations were performed on four clear nights (2015 June 17, 24, 26 and 27) using 0.36-m Cassegrain telescope with SBIG ST-8XME CCD. A total of 101 unfiltered images with 300-sec exposures were obtained (24, 19, 40 and 18 images covering 2.6, 2.0, 3.3 and 1.7 hours, respectively).

Using the combined data set resulting in 9.6 hours of effective time we have searched for the orbital period. The observing times were converted from JD to Barycentric Julian Date using online period search service http://scan.sai.msu.ru/lk/ by Kirill Sokolovsky. Using Lafler-Kinman and Deeming methods we have obtained the best value of period 0.06980(2) d, or 1.6752 hr. The nearby USNO-B1.0 1427-0327222 = SDSS J161945.69+524749.8 was used a reference star with R=16.15 and USNO-B1.0 1427-0327149 as a check star.

Then we have downloaded publicly available Catalina Sky Survey data (Drake et al., 2009) for this object from CRTS DR2 website http://nesssi.cacr.caltech.edu/DataRelease/ and combined them with our observations. CRTS has a total of 169 points for CSS_J161935.8+524631 from 2005 May 15 to 2013 October 24. Time of observations was converted from MJD to JD adding 2400000.5 to Catalina values and then corrected for barycenter, as described above. Applying Lafler-Kinman method to the combined dataset we have obtained the more precise value of period equal to 0.0697944(2) d.

The light curve of J1619+5246 from CRTS data during 2005 to 2013 folded with the best period is shown at Figure 3, and the light curve from our observations in 2015 is given at Figure 4. The change in the shape of the light curve is obvious. Before 2014 there was one prominent peak on the period near the zero phase, while the 2015 light curve is characterized by two peaks nearly equal in height. Also, the average level of brightness has increased in 2015 compared to 2005-2013 interval.

## DISCUSSION

X-ray source 1RXS J161935.7+524630 is characterized by hardness ratios HR1=–0.45±0.09, HR2=0.42±0.18 in ROSAT catalog. The negative value of HR1 differs remarkably from the typical numbers for other subtypes of cataclysmic variables, namely for dwarf novae (HR1 from 0.9 to 1.0). The values like that are actually more typical for quasars rather than for CVs. Combined with 2MASS (J-K) color index of 1.2 that has prevented this CV from discovery back in 2007. However, the value of (J-K) for the objects at J=16.7 and K=15.5 should be taken with caution since these infrared magnitudes are close to the limit of 2MASS survey.

| Variable name | 1RXS name | X-ray count rate | Hardness 1 | Hardness 2 |
|---|---|---|---|---|
| BY Cam | J054249.9+605132 | 1.499±0.066 | –0.10±0.04 | 0.47±0.05 |
| V1432 Aql | J194011.6–102529 | 0.318±0.033 | 0.72±0.08 | 0.23±0.11 |
| CD Ind | J211540.9-584045 | 0.380±0.044 | –0.02±0.10 | 0.39±0.14 |
| RX J0524+42 | J052430.2+424449 | 0.075±0.014 | 0.61±0.19 | 0.46±0.19 |
| DDE 32 | J161935.7+524630 | 0.135±0.017 | –0.45±0.09 | 0.42±0.18 |

Table 1. Comparison of X-ray parameters of selected known polars with asynchronously rotating white dwarfs and DDE 32.

The star is also an UV source GALEX J161935.8+524631 with the far and near UV magnitudes FUV=18.75±0.10, NUV=18.96±0.06. They are matching well with SDSS magnitudes (u=18.87 g=18.99 r=18.85 i=18.50 z=18.28). The optical (g-r), (r-i) indices correspond well to the color-period CV relation for an object with P=0.0698 d. Though the relationship by Kato et al. (2012) is based mostly on SDSS colors of dwarf novae, it can be used for estimating the orbital periods and spectral types of secondary components for the polar CVs. The contribution of the secondary component in this binary system is too small to cause the large (J-K) infrared excess in 2MASS catalog.

Pole switching has been observed in several magnetic cataclysmic binary systems before. The most known example is BY Cam (Mason et al., 1998; Pavlenko et al., 2007). This system is characterized by the WD rotation period close to the orbital one ($P_{spin}=0.990*P_{orb}$) which is causing beat phenomenon. Further study is required to measure the rotation period of WD in DDE 32 = J1619+5246 and to verify its classification as AM Her ("normal" polar) or BY Cam (asynchronous rotator) system.

## CONCLUSION

We conclude that DDE 32 = 1RXS J161935.7+524630 is a new magnetic cataclysmic variable (polar) with the orbital period P=0.0697944(2) days and changing geometry of accretion. The interval of 2005-2013 covered by Catalina Sky Survey was characterized by the accretion to one magnetic pole of the white dwarf, as seen from the single peak on the light curve (Figure 3). On the other hand, during our observations in June 2015 the object was in two-pole accretion mode, as shown by two peaks of nearly equal height (Figure 4). Since the phase of the first peak in 2015 data matches that of main peak in 2005-2013, we conclude that this peak is caused by the persistent hot spot near one of magnetic poles on WD surface. The second peak near the phase 0.5 is the second hot spot on the other magnetic pole that has "turned on" at some moment in Spring 2015, as seen from the increasing level of brightness in 2014-2015 BRT data. There is a possibility of DDE 32 being an asynchronous rotator similar to BY Cam. The continued monitoring of this exciting cataclysmic binary system is highly encouraged. The time-resolved spectroscopy and spectral lines tomography is very much desired for this non-stationary object in Draco.

FIGURES

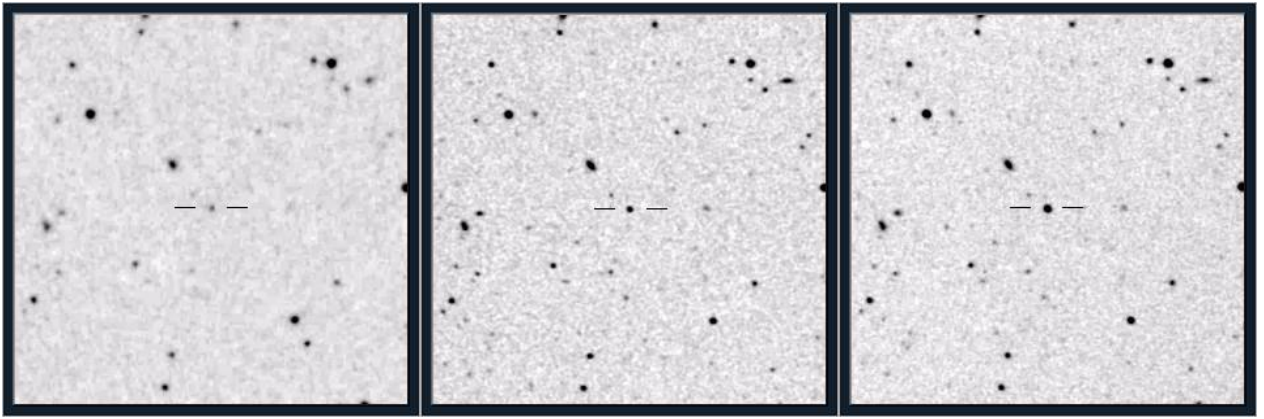

Figure 1. DDE 32 = 1RXS J161935.7+524630 on the digitized Palomar Red plates. Left to right: 1954 June 28 (faint state), 1991 July 03 (medium brightness) and 1993 April 25 (bright state). FOV is 5'x5', North is up and East is at left.

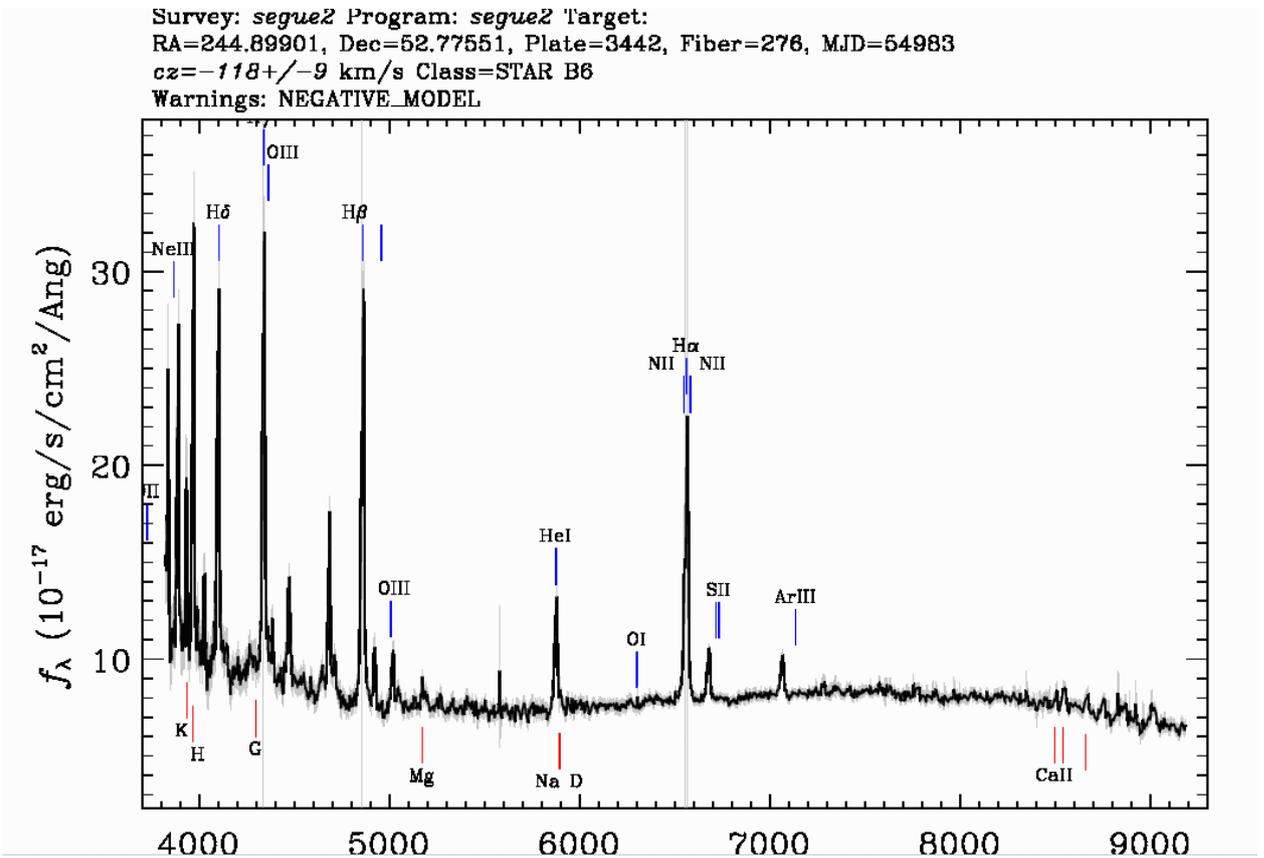

Figure 2. SDSS spectrum of 1RXS J161935.7+524630 taken on 2009 June 01.

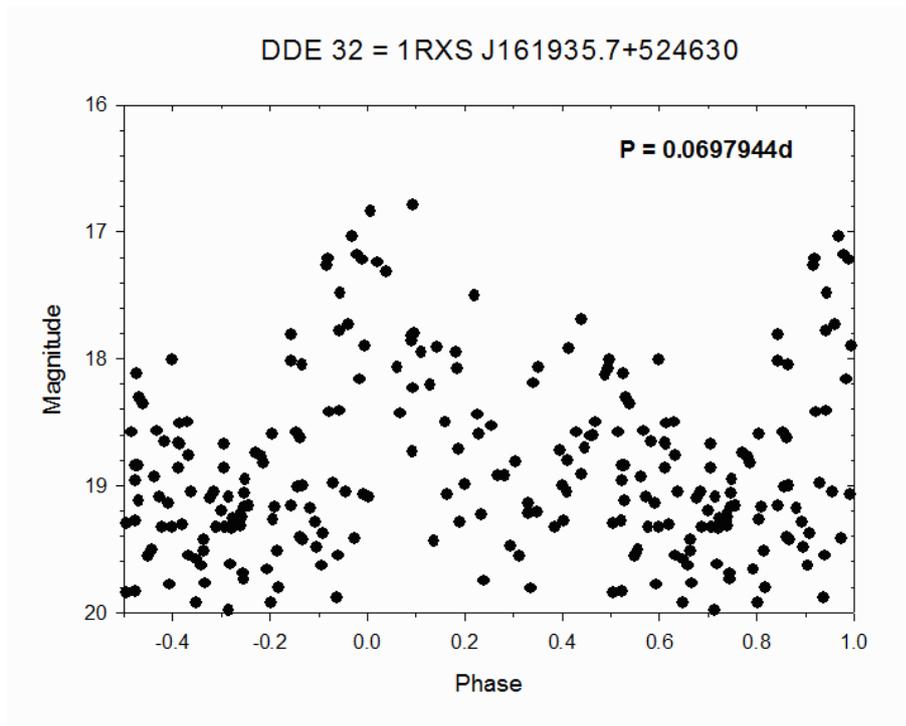

Figure 3. Light curve of J1619+5246 from Catalina Sky Survey data (2005-2013) folded with the orbital period P=0.0697944 d.

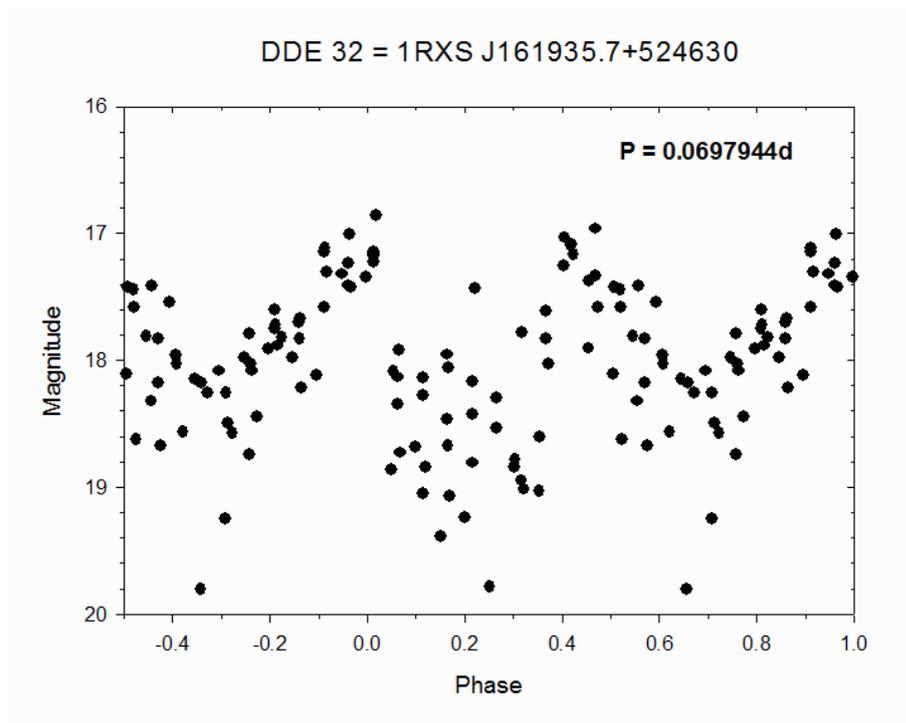

Figure 4. Light curve of J1619+5246 from Lajatico data (June 2015) folded with the same orbital period and the same initial epoch $T_0$=2454923.966 (HJD).